\documentclass{article}

\usepackage{arxiv}

\usepackage{amssymb}
\usepackage{amsmath}
\usepackage[utf8]{inputenc} 
\usepackage[T1]{fontenc}    
\usepackage{hyperref}       
\usepackage{url}            
\usepackage{booktabs}       
\usepackage{amsfonts}       
\usepackage{nicefrac}       
\usepackage{microtype}      
\usepackage{xcolor}         
\usepackage{algorithm}
\usepackage{algpseudocode}
\usepackage{graphicx}
\usepackage{multirow}
\usepackage[flushleft]{threeparttable}
\usepackage{float}
\usepackage{caption}
\usepackage{mathrsfs}

\title{GOLFer: Smaller LM-Generated Documents Hallucination Filter \& Combiner for Query Expansion in Information Retrieval}

\author{
 Lingyuan Liu \\
  City University of Hong Kong \\
  \texttt{ly.liu@my.cityu.edu.hk} \\
   \And
 Mengxiang Zhang\footnote{$*$} \\
  The University of Hong Kong \\
  \texttt{mxzhang6@connect.hku.hk} \\
}

\begin{document}
\maketitle
\begin{abstract}
Large language models (LLMs)-based query expansion for information retrieval augments queries with generated hypothetical documents with LLMs. However, its performance relies heavily on the scale of the language models (LMs), necessitating larger, more advanced LLMs. This approach is costly, computationally intensive, and often has limited accessibility. To address these limitations, we introduce GOLFer - Smaller LMs-\underline{G}enerated D\underline{o}cuments Ha\underline{l}lucination \underline{F}ilter \& Combin\underline{er} - a novel method leveraging smaller open-source LMs for query expansion. GOLFer comprises two modules: a hallucination filter and a documents combiner. The former detects and removes non-factual and inconsistent sentences in generated documents, a common issue with smaller LMs, while the latter combines the filtered content with the query using a weight vector to balance their influence. We evaluate GOLFer alongside dominant LLM-based query expansion methods on three web search and ten low-resource datasets. Experimental results demonstrate that GOLFer consistently outperforms other methods using smaller LMs, and maintains competitive performance against methods using large-size LLMs, demonstrating its effectiveness.
\footnotetext{Corresponding author.}
\footnote{The code for our method is publicly available at \href{https://github.com/liuliuyuan6/GOLFer}{https://github.com/liuliuyuan6/GOLFer}.}

\end{abstract}

\section{Introduction}\label{sec:01}
Information retrieval (IR) is crucial for extracting relevant information from large repositories, serving as a key component in modern search engines \cite{wang2019multi, karpukhin2020dense}. Query expansion, a key technique for enhancing IR performance, improves the precision and expressiveness of user queries \cite{azad2019query}. Traditional methods use hand-built knowledge resources like WordNet and Thesaurus (\cite{pal2014improving, gong2005web} or external text collections \cite{roy2016using, diaz2016query}. However, these methods are limited by the quality of external data sources and show limited success on popular datasets \cite{azad2019query}. More adaptable query expansion approaches are needed to meet diverse requirements across different contexts.

\begin{figure*}[t]
\centering
 \includegraphics[scale = 0.7]{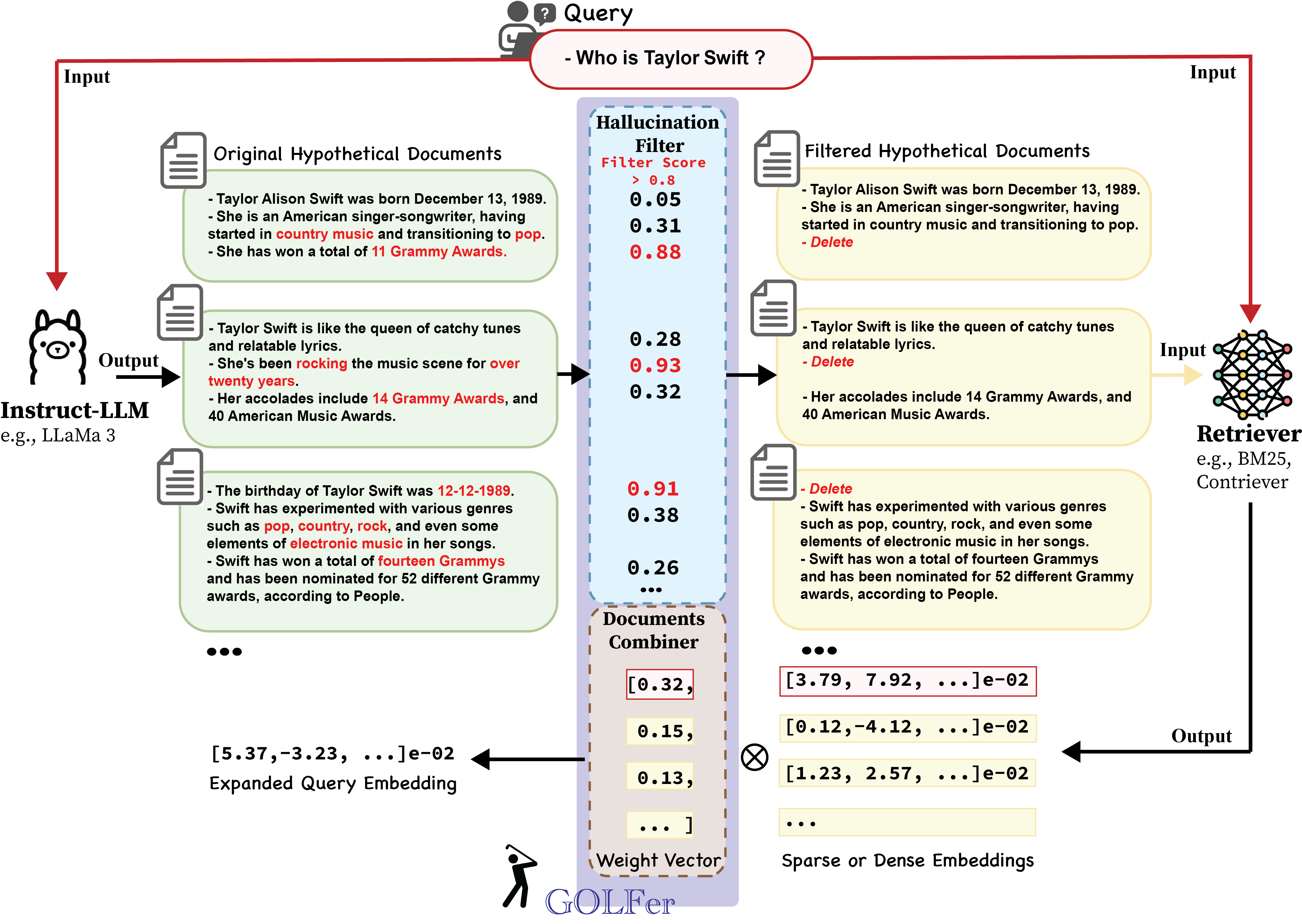} 
 \caption{\small Overview of GOLFer. Given a query, GOLFer generates $n$ passages using an Instruct-smaller LM, which are then processed through a hallucination filter to produce filtered hypothetical documents. These filtered documents are combined with the original query using a weight vector by the documents combiner module to create the expanded query embedding for retrieval.}\label{fig:GOLFer}
\end{figure*}

LLMs like GPT-4 \cite{ouyang2022training} and LLaMA 3 \cite{dubey2024llama} have shown impressive abilities in generating fluent and realistic responses. Pre-trained on extensive corpora, these models excel in natural language understanding and generation. \cite{ouyang2022training} indicates that LLMs can be fine-tuned with minimal data to align with human intent, enabling them to generalize to diverse instructions in a zero-shot manner. This adaptability has spurred interest in using LLMs for query expansion in IR, where queries are often brief or ambiguous \cite{mitra2017neural, zhao2024dense}. LLMs can generate hypothetical documents based on various prompts to enhance query expansion. For example, HyDE uses zero-shot instructions to generate a hypothetical document \cite{gao2022precise}, and Query2Doc employs few-shot examples to create hypothetical-documents \cite{wang2023query2doc}. These methods enhance the performance of retrievers such as Contriever and BM25 across a variety of tasks, including web search, question answering, and fact verification.

However, existing LLM-based query expansion methods face several critical challenges. Empirical experiments, such as those involving HyDE, indicate that the performance of LLM-based query expansion heavily depends on the scale of the LLM employed \cite{gao2022precise}. \cite{wang2023query2doc} suggest that smaller LMs tend to produce shorter outputs with more factual errors, posing a significant obstacle to building trustworthy systems. This tendency to hallucinate facts has been widely observed, where models can confidently generate fictitious information. Consequently, current LLM-based query expansion methods often advocate for using large-scale, advanced LLMs like GPT-3.5 (175B) and GPT-4 to mitigate these inaccuracies and enhance query expansion performance in IR. However, in practice, employing such large-scale models is costly, computationally intensive, and regionally restricted. For instance, the API costs for GPT-3.5 and GPT-4 are \$1.50 and \$2.50 per million tokens, respectively \footnote{Refer to \url{https://openai.com/api/pricing/}.}, making widespread use financially prohibitive for small and private institutions. Additionally, quota limits often restrict the practical application of these models, and their token-by-token autoregressive decoding can significantly reduce retrieval efficiency, potentially taking over 2000 ms to generate hypothetical documents \cite{wang2023query2doc}. Furthermore, access to large-scale LLMs like GPT is restricted in certain regions, such as China, further limiting their applicability. These factors collectively constrain the practical deployment of existing LLM-based query expansion methods.

To overcome these limitations, we introduce a new method, GOLFer, i.e., smaller LMs-\underline{g}enerated d\underline{o}cuments ha\underline{l}lucination \underline{f}ilter \& combin\underline{er} for query expansion in IR (See Fig. \ref{fig:GOLFer}). Unlike existing LLM-based query expansion methods, our method focuses on smaller open-source LMs, such as LLaMA-3-8B-Instruct, to assist in query expansion, thereby mitigating some of the limitations associated with using large-scale models. However, smaller LMs often introduce hallucinations, such as non-factual content and inconsistent contexts, when generating hypothetical documents. If these flawed documents are directly utilized by retrievers, they can introduce irrelevant, noisy, or erroneous data, jeopardizing the quality of the expanded query embeddings and thus affecting query expansion performance in IR. To address this issue, GOLFer is designed to detect and filter out non-factual and inconsistent sentences from the hypothetical documents, ensuring that only relevant and accurate information is incorporated. Additionally, to balance the influence of the query and hypothetical documents, our method combines these filtered documents with the original query based on factuality by setting a weight vector. The inner product of the weight vector and the embedding vectors for the original query and filtered documents forms a new, refined query for IR. GOLFer is a lightweight, versatile query expansion method that can be integrated into any Transformer-based LM without requiring additional training, fine-tuning, or prompt engineering.

We comprehensively evaluate GOLFer using LLaMA-3-8B-Instruct as the smaller LM, alongside three types of retrievers: sparse retriever (e.g., BM25), dense retriever (e.g., ANCE \cite{xiong2020approximate}), and advanced dense retriever (e.g., Aggretriever\_v1 \cite{lin2023aggretriever}). Under consistent experimental conditions, we compare GOLFer with existing LLM-based query expansion methods, including HyDE \cite{gao2022precise} and query2doc \cite{wang2023query2doc}, across three web search datasets from MS MARCO \cite{bajaj2016ms} and ten low-resource datasets from the BEIR benchmark \cite{thakur2021beir}. The results demonstrate that GOLFer outperforms other methods using smaller LMs on all datasets across nearly all evaluation metrics, highlighting its effectiveness. Notably, GOLFer remains competitive even against methods utilizing large-scale LLMs like GPT-3.5 (175B). 

In summary, the contributions of our paper are as follows:
\begin{itemize}
  \item We propose a novel smaller LMs-driven query expansion method: GOLFer. It can i) detect and filter out non-factual and inconsistent sentences in the original hypothetical documents generated by smaller LMs, and ii) combine these filtered hypothetical documents with the original query by a weight vector to form a expanded query for IR, enhancing factuality. 
  \item We evaluate GOLFer alongside dominant query expansion methods on MS MARCO dev, TREC DL19 and DL20 and nine low-resource datasets form BEIR. Experimental results demonstrate the efficacy of our method.
\end{itemize}

\section{Related Work}\label{sec:02}

\subsection{LLM-based Query Expansion}\label{subsec:0201}
Query expansion improves retrieval systems by broadening query terms to include synonyms or related concepts, enhancing document matching. Leveraging LLMs' generative capabilities, some studies generate hypothetical documents for query expansion. For instance, HyDE uses an instruction-following LLM to create a hypothetical document \cite{gao2022precise}, which is then encoded into an embedding vector for retrieval. Similarly, Query2Doc generates hypothetical documents through few-shot prompting of LLMs \cite{wang2023query2doc}, combining them with the original query to boost retrieval performance. Additionally, various prompting strategies to generate hypothetical documents for query expansion, including zero-shot, few-shot, and Chain-of-Thought, have been explored \cite{jagerman2023query}, with Chain-of-Thought particularly effective. These advances suggest that knowledge distillation from LLMs can transfer their capabilities to smaller models, advancing LLM-based query expansion.

\subsection{Hallucination Detection in LLM-Generated Documents}\label{subsec:0202}
LLM-generated documents often suffer from hallucinations, producing nonsensical or inaccurate text \cite{raunak2021curious}, which degrades system performance \cite{welleck2019neural}. Recent research has focused on identifying these hallucinations through three primary approaches: white-box, grey-box, and black-box methods. White-box methods leverage LLM internal states to assess response factuality \cite{azaria2023internal}, requiring labeled data for supervised training. Grey-box methods evaluate factuality using output distributions, employing intrinsic uncertainty metrics to identify uncertain segments \cite{yuan2021bartscore, fu2023gptscore}.  Black-box methods, like SelfCheckGPT \cite{manakul2023selfcheckgpt}, fact-check responses by comparing multiple sampled outputs for consistency. Each approach addresses hallucinations in different contexts, enhancing the reliability of LLM-generated content.

\section{Methodology }\label{sec:03}
In this section, we provide a detailed explanation of the GOLFer method. GOLFer is composed of two main components: the hallucination filter and the documents combiner, as shown in Fig. \ref{fig:GOLFer}. Within the GOLFer method, $n$ passages generated by an Instruct-smaller LLM from a given user query are treated as original hypothetical documents for query expansion. Initially, these documents are processed through the hallucination filter to produce filtered hypothetical documents. Subsequently, the expanded query embedding for query expansion is obtained by combining embedding vectors with a weight vector, which is determined by the documents combiner module in the GOLFer method. The configuration of the weight vector varies between sparse and dense retrieval methods. The hallucination filter is detailed in section \ref{subsec:0301}, while the documents combiner is discussed in section \ref{subsec:0302}.

\paragraph{Notation}

Suppose $n$ passages for query expansion are generated by an instruction-following smaller LM, such as LLaMa-3-8B-Instruct, given a user query $q$. Let $d^i$ refer to the $i$-th smaller LM-generated document, where $i \in \{1, 2, ..., n \}$. Each document $d^i$ contains $m_i$ sentences, denoted by $s^i_j$ for the $j$-th sentence in document $d^i$, where $j \in \{1, 2, ..., m_i \}$. Furthermore, each sentence $s^i_j$ consists of $o^i_j$ tokens, with $t^i_j(l)$ representing the $l$-th token in the $j$-th sentence of the $i$-th smaller LM-generated document, where $l \in \{1, 2, ..., o^i_j \}$.

\subsection{Hallucination Filter}\label{subsec:0301}
The hallucination filter module can evaluate the degree of hallucination for each sentence in a smaller LM-generated passages based on consistency and factuality, and then, can filter these sentences out based on their hallucination degree. 

Hallucination degree is based on the idea that factual sentences generated by a smaller LLM tend to contain tokens with higher likelihood and lower entropy, whereas hallucinated sentences are characterized by tokens with flat probability distributions and high uncertainty. What's more, each token has a different influence on the subsequent context. Thus, we define the factuality score of the hypothesis documents by evaluates the uncertainty of tokens and their impact on subsequent tokens. Our method begins by quantifying the uncertainty of each token, $t^i_j(l)$. This is achieved by recording the entropy of the token's probability distribution across the vocabulary. For any token $t^i_j(l)$, the entropy $\mathcal{H}^i_{j_l}$ is computed as follows: 

\begin{equation}\small\label{eq:03}
\mathcal{H}^i_{j_l} = - \sum_{\tilde{v} \in \mathcal{V}} p^i_{j_l}(\tilde{v})\log{p^i_{j_l}(\tilde{v})},
\end{equation}where $p^i_{j_l}(\tilde{v})$ denotes the probability of generating the token $\tilde{v}$ over all tokens in the vocabulary $\mathcal{V}$ at position $l$ of the $j$-th sentence in document $i$.

In addition to uncertainty, GOLFer leverages the self-attention mechanism inherent in Transformer-based LLMs to assign weights to tokens, reflecting their impact on the subsequent context. Specifically, for any given token $t^i_j(l)$, we quantify its influence by recording the average attention value $Avg(\mathcal{A}^i_{j_l})$, which captures the average attention from all following tokens. The attention scores are taken from the last Transformer layer of the smaller LM. The attention value $\mathcal{A}^i_{j_{l,v}}$ between two tokens $t^i_j(l)$ and $t^i_j(v)$ for any $l < v$ is computed as follows: 

\begin{equation}\small\label{eq:04}
\mathcal{A}^i_{j_{l,v}}= \textup{softmax} \left( \frac{Q^{i}_{j_l}{K^{i}_{j_v}}^\top }{ \sqrt{d_{k} }} \right),
\end{equation} where $Q^{i}_{j_l}$ represents the query vector of token $t^i_j(l)$, $K^{i}_{j_v}$ is the key vector of token $t^i_j(v)$, and $d_{k}$ denotes the dimensionality of the key vector. 
The softmax function is applied to the dot product of $Q^{i}_{j_l}$ and $K^{i}_{j_v}$, normalized by the square root of $d_{k}$. The average attention value $Avg(\mathcal{A}^i_{j_l})$ for token $t^i_j(l)$ is then identified by averaging $\mathcal{A}^i_{j_{l,v}}$ for all $v > l$:

\begin{equation}\small\label{eq:05}
Avg(\mathcal{A}^i_{j_l})= \frac{\sum_{v = l+1}^{o^i_j} \mathcal{A}^i_{j_{l,v}}}{o^i_j-l} .
\end{equation} Combining uncertainty and significance, GOLFer computes a comprehensive factuality score for each token $t^i_j(l)$. Specifically, the factuality score $\mathscr{F}^i_{j_l}$ is calculated by multiplying the entropy $\mathcal{H}^i_{j_l}$ of the token by its average attention value $Avg(\mathcal{A}^i_{j_l})$: 

\begin{equation}\small\label{eq:06}
\mathscr{F}^i_{j_l} = \mathcal{H}^i_{j_l}\cdot Avg(\mathcal{A}^i_{j_l}). 
\end{equation} This token-level factuality score serves as the basis for evaluating the overall factuality of a sentence. The sentence-level factuality score $\mathscr{F}(s^i_j)$ is then derived by averaging the factuality scores of all tokens within the sentence: 

\begin{equation}\small\label{eq:07}
\mathscr{F}(s^i_j) = \frac{\sum_{l=1}^{o^i_j}\mathscr{F}^i_{j_l}}{o^i_j}. 
\end{equation} 

And the fundamental idea behind detecting hallucination in terms of consistency is rooted in the premise that if a smaller LLM possesses genuine knowledge of a concept, its sampled responses will likely be similar and factually consistent. Conversely, hallucinated facts often lead to divergent and contradictory responses when multiple outputs are drawn from the same query. By comparing multiple responses generated from the same query, we can assess information consistency and determine the factuality of statements \cite{manakul2023selfcheckgpt}. Natural Language Inference (NLI) has been utilized to measure faithfulness in hallucination detection \cite{manakul2023selfcheckgpt}, demonstrating competitive performance. Inspired by this approach, we employ a fine-tuned NLI classifier, DeBERTa-v3-large \cite{he2021debertav3}, to compute the NLI contradiction score for each sentence across different documents. Only the logits associated with the entailment and contradiction classes are considered, and the NLI contradiction score $\mathscr{C}$ for the sentence $s^i_j$ in document $d^{k \neq i}$ is computed as follows:

\begin{equation}\small\label{eq:01}
\mathscr{C}(s^i_j, d^{k\neq i}) = \frac{\exp(\omega_c)}{\exp(\omega_c)+\exp(\omega_e)},
\end{equation} where $\exp(\omega_c)$ and $\exp(\omega_e)$ are the logits of the contradiction and entailment classes, respectively. In GOLFer, the consistency score for a sentence $s^i_j$ is the mean value of its NLI contradiction scores across all documents that do not contain $s^i_j$. This can be formulated as:

\begin{equation}\small\label{eq:02}
\mathscr{C}(s^i_j) = \frac{\sum_{k\in\{1,2,...,n\}\backslash\{i\}}  \mathscr{C}(s^i_j, d^{k})}{n-1}. 
\end{equation} We calculate the filter score $\mathscr{H}(s^i_j)$ as follows:
\begin{equation}\small\label{eq:0201}
\mathscr{H}(s^i_j) = \mathscr{F}(s^i_j) \cdot \mathscr{C}(s^i_j). 
\end{equation} 
If the filter score $\mathscr{H}(s^i_j)$ is beyond than a certain number, we will delete it since it could be highly hallucinatory. We have empirically found that 0.8 is a generally good value and do not tune it on a dataset basis.

\subsection{Documents Combiner}\label{subsec:0302}
The documents combiner module can merge the original query with filtered hypothetical documents to form an expanded query for IR. This combination process is tailored based on the type of retrieval—sparse or dense—and the generation confidence of the documents. The subsequent sections elaborate on the specific operations for sparse and dense retrievers, respectively.

\paragraph{Sparse Retrieval} 
In the case of sparse retrieval, we enhance the query term weights by repeating the query 20 times when combining 5 hypothesis documents. This repetition aims to balance the relative weights of the query and the hypothetical documents before merging them. The expanded query $q^+$ for sparse retrieval is then formulated as follows: 
\begin{equation}\small\label{eq:11}
q^{+}=20 \cdot q + \sum_{i=1}^{n}d^i,
\end{equation}where $n=5$.
This formulation ensures an effective balance between the original query and the augmented content for improved retrieval performance.

\paragraph{Dense Retrieval}
For dense retrieval, the document combiner module takes into account the generation confidence of smaller LM-generated hypothetical documents. These documents, with varying degrees of generation confidence, exhibit different levels of factual information and relevance patterns concerning the real documents we aim to retrieve. Consequently, we posit that hypothetical documents with higher generation confidence should contribute more significantly to the query expansion process for IR.

Specifically, we estimate the generation confidence, $\varpi(d^i)$ of a filtered document by averaging the generation probabilities, $p^i_{j_l}$, of each token within the filtered documents. This can be expressed as follows: 

\begin{equation}\small\label{eq:08}
\varpi(d^i) = \frac{\sum^{m_i}_{j=1}\sum^{o^i_j}_{l=1}p^i_{j_l}}{\sum^{m_i}_{j=1} o^i_j}. 
\end{equation} Subsequently, we encode both the generated documents and the original query into embedding vectors using the dense retriever, denoted as $f(\cdot)$. The expanded query embedding for IR, $V_{q^+}$, is then formulated by combining the embedding vectors for the filtered hypothetical documents and the original query with their corresponding weights. This can be expressed as follows: 

\begin{equation}\small\label{eq:10}
V_{q^+}=\beta \cdot f(q)+\frac{1-\beta}{\sum^{n}_{i=1}\varpi(d^i)} \cdot \sum^{n}_{i=1} \varpi(d^i)f(d^i) ,
\end{equation}
where $\beta$ represents the contribution rate of the original query in forming the expanded query for IR. And we find that $\beta=0.6$ is an effective values for combing 5 hypothesis documents , which we do not tune on a dataset-specific basis.  For dense retrieval, the inner product is computed between $V_{q+}$ and the set of all document vectors, and the most similar documents are subsequently retrieved.

\section{Experiments }\label{sec:04}
\begin{table*}[!ht]\tiny
    \centering
    \begin{tabular}{lc|ll|lll|lll}
    \hline
        \multirow{2}{*}{Model} & \multirow{2}{*}{Fine-tuning}  &  \multicolumn{2}{c}{\textbf{MS MARCO dev} } &  \multicolumn{3}{c}{\textbf{TREC DL 19} } & \multicolumn{3}{c}{\textbf{TREC DL 20} } \\ 
        &  & MRR@10 &  R@1k & MAP & nDCG@10 & R@1K & MAP & nDCG@10 & R@1K \\ \hline
        \multicolumn{10}{l}{\textit{Sparse retrieval} }  \\
        BM25 \cite{robertson2009probabilistic} & $\times$ & 18.4  & 85.7  & 30.1  & 50.6  & 75.0  & 28.6  & 48.0  & 78.6  \\ 
        ~~~~~~+query2doc \cite{wang2023query2doc} & $\times$ & 19.0 & 86.9 & 36.3 & 53.9 & 78.7 & 38.4 & 56.1 & 83.6 \\ 
        ~~~~~~+\textbf{GOLFer} & $\times$ & \textbf{19.9$^{+1.5}$} & \textbf{88.5$^{+2.8}$} & \textbf{39.1$^{+9.0}$} & \textbf{59.5 $^{+9.0}$} & \textbf{83.0$^{+7.9}$} & \textbf{45.5$^{+17.0}$} & \textbf{62.9$^{+14.9}$} & \textbf{86.1$^{+7.4}$} \\ \hline
        \multicolumn{10}{l}{\textit{Dense retrieval w/o distillation } }  \\
        ANCE \cite{xiong2020approximate} & $\surd$ & 33.0  & 95.9  & 37.1  & 64.5  & 75.5  & 40.8  & 64.6  & 77.6  \\ 
        ~~~~~~+HyDE \cite{gao2022precise} & $\surd$ & 33.2 & 96.3 & \textbf{45.3}  & 68.2  & \textbf{80.5}  & 44.2  & 67.8  & 81.6  \\ 
        ~~~~~~+query2doc \cite{wang2023query2doc} & $\surd$ & 32.9 & 96.0 & 45.0  & 69.9  & 77.8  & 44.6  & 67.5  & \textbf{82.0}  \\ 
        ~~~~~~+\textbf{GOLFer} & $\surd$ & \textbf{33.3$^{+0.3}$} & \textbf{96.4$^{+0.5}$} & 44.1$^{+7.0}$ & \textbf{ 71.3$^{+6.8}$} & 79.1$^{+3.6}$ & \textbf{44.7$^{+3.9}$} & \textbf{67.9$^{+3.3}$} & 81.2$^{+3.6}$ \\ \hline
        \multicolumn{10}{l}{\textit{Dense retrieval w/ distillation } }  \\        
         Colbert\_v2 \cite{DBLP:journals/corr/abs-2112-01488} & $\surd$ & 34.4  & 96.7  & 41.0  & 68.4  & 81.3  & 45.2  & 69.3  & 83.9  \\ 
        ~~~~~~+HyDE \cite{gao2022precise} & $\surd$ & 34.1 & 96.7 & 47.5  & 72.0  & 83.7  & 46.8  & 69.5  & 84.4  \\   				
        ~~~~~~+query2doc \cite{wang2023query2doc} & $\surd$ & 33.9 & 96.6 & 46.0  & 71.3  & 81.3  & 47.4  & 69.8  & 85.6  \\   
        ~~~~~~+\textbf{GOLFer} & $\surd$ & 34.5$^{+0.1}$ & 97.1$^{+0.4}$ & 48.2$^{+7.3}$ & 73.4$^{+4.9}$ & 84.6$^{+3.3}$ & 47.4$^{+2.2}$ & 70.0$^{+0.7}$ & 85.7$^{+1.8}$ \\ 
        Aggretriever\_v1 \cite{lin2023aggretriever} & $\surd$ & 34.1  & 96.0  & 43.0  & 68.2  & 80.2  & 43.3  & 67.3  & 83.5  \\ 
        ~~~~~~+HyDE \cite{gao2022precise} & $\surd$ & 33.8 & 96.3 & 47.3  & 68.9  & 85.7  & 44.8  & 68.0  & 84.4  \\ 				
        ~~~~~~+query2doc \cite{wang2023query2doc} & $\surd$ & 34.0 & 96.2 & 47.5  & 69.5  & 81.9  & 45.5  & 68.1  & 87.0  \\   
        ~~~~~~+\textbf{GOLFer} & $\surd$ & 34.3$^{+0.2}$ & 96.9$^{+0.8}$ & 48.3$^{+5.3}$ & 70.3$^{+2.1}$ & 85.3$^{+5.1}$ & 45.3$^{+2.0}$ & 67.9$^{+0.7}$ & 86.6$^{+3.1}$ \\ 
        Aggretriever\_v2 \cite{lin2023aggretriever} & $\surd$ & 36.2  & 97.4  & 43.5  & 68.4  & 80.8  & 47.1  & 69.7  & 85.6  \\   
        ~~~~~~+HyDE \cite{gao2022precise} & $\surd$ & 36.1 & 97.5 & 48.3  & \textbf{73.5}  & 85.4  & \textbf{49.2}  & 72.0  & 86.6  \\   
        ~~~~~~+query2doc \cite{wang2023query2doc} & $\surd$ & 36.1 & 97.4 & 46.4  & 72.1  & 81.8  & 47.5  & 71.1  & 88.0  \\  
        ~~~~~~+\textbf{GOLFer} & $\surd$ & \textbf{36.3$^{+0.1}$} & \textbf{97.8$^{+0.4}$} & \textbf{48.4$^{+4.9}$} & 73.0$^{+4.7}$ & \textbf{86.4$^{+5.6}$} & 49.0$^{+1.9}$ & \textbf{72.2$^{+2.5}$} & \textbf{88.3$^{+2.8}$} \\ \hline
    \end{tabular}
        \caption{Results for web search on MS MARCO dev and DL19/20. Best performing systems are marked \textbf{bold}. All the hypothetical documents used in this table are generated by LLaMA-3-8B-Instruct.}
    \label{tab1}
\end{table*}

\begin{table*}[!ht]\scriptsize
    \centering
    \begin{tabular}{lccc|cccc} \hline
         \multirow{2}{*}{Dataset} & \multicolumn{7}{c}{nDCG@10} \\ 
         & BM25 & BM25+G. & BM25+Q. & Cont. & Cont.+G. & Cont.+H. & Cont.+Q. \\ \cline{2-8}
        \textbf{NQ}  & 30.5  & \textbf{47.6}  &  42.3 & 49.8  & \textbf{51.2}  &  51.1 &  50.8 \\ 
        \textbf{FiQA-2018}  & 23.6  & \textbf{23.9}  & 23.6 & 24.5  & \textbf{26.5}  & 24.5 & 21.3 \\ 
        \textbf{TREC-COVID}  & 59.5  & \textbf{69.9}  &  72.1 & 27.1  & \textbf{57.4}  &  53.1 &  51.5 \\ 
        \textbf{Signal-1M}  & 33.0  & \textbf{36.5}  &  34.7 & 27.8  & \textbf{29.9}  &  29.1 &  29.4 \\ 
        \textbf{TREC-NEWS}  & 39.5  & \textbf{50.5}  & 49.1 & 34.8  & \textbf{41.1}  & 40.1 & 38.7 \\ 
        \textbf{Robust04}  & 40.7  & \textbf{46.8}  &  43.1 & 47.3  & \textbf{47.7}  &  47.5 &  47.4 \\
        \textbf{Touche 2020}  & 44.2  & \textbf{45.8}  & 45.3 & 20.4  & 21.1  & 20.7 & \textbf{21.8} \\  
        \textbf{CQADupStack}  & 30.2  & \textbf{31.4}  &  30.0 & \textbf{34.5}  & 34.4  &  34.2 &  33.9 \\  
        \textbf{DBPedia}  & 31.3  & \textbf{34.1}  & 32.8 & 41.3  & \textbf{46.5}  & 42.3 & 43.2 \\  
        \textbf{SciFact}  & 67.9  & \textbf{70.8}  & 70.3 & 67.7  & \textbf{69.4}  & 67.9 & 66.2 \\ \hline
        ~ & \multicolumn{7}{c}{Recall@100}  \\  
        \textbf{NQ}  & 76.0  & \textbf{89.7}  &  85.3 & 82.1  & \textbf{83.1}  &  82.9 &  82.8 \\  
        \textbf{FiQA-2018}  & 53.9  & \textbf{56.9}  & 56.0 & 56.2  & \textbf{61.8}  & 59.3 & 56.8 \\  
        \textbf{TREC-COVID}   & 49.8  & \textbf{56.7}  &  51.5 & 17.2  & \textbf{32.0}  & 30.4  & 30.4  \\  
        \textbf{Signal-1M}  & 37.0  & \textbf{39.8}  &  38.1 & 32.2  & \textbf{34.4}  &  33.6 &  33.1 \\  
        \textbf{TREC-NEWS}  & 44.7  & \textbf{52.8}  & 51.5 & 42.3  & \textbf{49.7}  & 46.1 & 43.2 \\  
        \textbf{Robust04}  & 37.5  & \textbf{38.3}  &  37.7 & 39.2  & \textbf{42.6}  &  40.3 &  41.1 \\  
        \textbf{Touche 2020}  & 53.8  & 56.7  &  \textbf{56.8}& 44.2  & \textbf{46.0}  &  44.4 & 45.8 \\  
        \textbf{CQADupStack} & 60.6  & \textbf{61.7}  &  60.5 & \textbf{66.3}  & 59.7  &  59.1 &  58.7 \\  
        \textbf{DBPedia} & 39.8  & \textbf{47.0}  & 42.1 & 54.1  & \textbf{58.1}  & 46.2 &  45.1 \\  
        \textbf{SciFact} & 92.5  & \textbf{95.4}  & 95.1 & 92.6  & \textbf{96.6}  & 96.1 & 94.1  \\ \hline
    \end{tabular}
     \caption{Results for Low resource tasks from BEIR. Best performing systems are marked \textbf{bold}. G. represents GOLFer, Q. represents query2doc, and H. represents HyDE, and Cont. represents Contriever that are fine-tuned on MS MARCO training data. All the hypothetical documents used in this table are generated by LLaMA-3-8B-Instruct.}
    \label{tab2}
\end{table*}

\subsection{Setup} \label{subsec:0401}
\paragraph{Implementation}
We implement Meta-LLaMA-3-8B-Instruct \cite{dubey2024llama}, a smaller open-source LM, to generate hypothetical documents for given queries. We sample documents with a temperature setting of 0.6, top-p of 0.9 and max tokens of 128 for open-ended generation. Retrieval experiments are conducted using the Pyserini toolkit \cite{lin2021pyserini}.

\paragraph{Datasets and Evaluations}
We evaluate our method using two types of datasets relevant to information retrieval tasks. The first type includes web search datasets: MS MARCO dev \cite{bajaj2016ms}, TREC-DL-2019  \cite{craswell2020overview}, and 2020 \cite{craswell2021overviewtrec2020deep}. 
The second type consists of low-resource datasets from the BEIR benchmark \cite{thakur2021beir}, such as NQ, FiQA-2018, TREC-COVID, Signal-1M, TREC-NEWS, Robust04, Touche2020, CQADupStack, DBPedia, and Scifact. We use the following evaluation metrics: $MAP$, $nDCG@10$, and $Recall@1k$ for TREC DL 2019 and 2020, $MRR@10$ and $Recall@1k$ for MS-MARCO datasets, and $nDCG@10$ and $Recall@100$ for the BEIR datasets.
We employ distinct instructions for each dataset, maintaining a similar structure but varying quantifiers to control the form of the generated hypothetical documents. These instructions are detailed in Appendix \ref{A:0101}.

\paragraph{Compared Systems}
In our experiments, for the web search task, we use BM25 \cite{robertson2009probabilistic} as the baseline for sparse retrieval and ANCE \cite{xiong2020approximate}, fine-tuned on MS-MARCO datasets, as the baseline for dense retrieval. Additionally, we consider three advanced dense retrievers enhanced by distillation and pre-training techniques: Colbert\_v2 \cite{DBLP:journals/corr/abs-2112-01488}, Aggretriever\_v1 trained with distillation \cite{lin2023aggretriever}, and Aggretriever\_v2 trained with distillation and pre-training \cite{gao2021unsupervised}.
For the low-resource retrieval task, BM25 is again used as the baseline for sparse retrieval, while Contriever \cite{izacard2021towards} serves as the baseline for dense retrieval. Retrievers within GOLFer share the same embedding spaces as these baselines, with the primary difference being in how the query vector is constructed.

This setup allows us to effectively assess the impact of GOLFer. Furthermore, we compare our method with two other query expansion approaches: HyDE, designed for dense retrieval systems, and query2doc, applicable to both sparse and dense retrieval systems. HyDE and query2doc use the same original hypothesis documents as we do.

\subsection{Web Search} \label{subsec:0402}
The results, summarized in Table \ref{tab1}, present the performances of various retrieval models enhanced by GOLFer. For sparse retrieval, GOLFer consistently outperforms the query2doc approach across all metrics, demonstrating its superior effectiveness in improving retrieval performance. In dense retrieval without distillation, ANCE combined with GOLFer shows significant improvements across most metrics compared to other methods. For dense retrieval with distillation, GOLFer enhances the performance of Colbert\_v2, Aggretriever\_v1 and v2, with notable gains in metrics such as $nDCG@10$ and $R@1k$. This consistent improvement across both sparse and dense retrieval models highlights the robustness and reliability of GOLFer in enhancing retrieval systems.

\subsection{Low Resource Retrieval} \label{subsec:0403}
The performance for low-resource tasks is summarized in Table \ref{tab2}. For sparse retrieval using BM25, GOLFer generally outperforms query2doc across nearly all datasets. Notably, GOLFer improves $nDCG@10$ and $Recall@100$ significantly on datasets such as NQ, TREC-COVID, and TREC-NEWS. The only exception is a slight underperformance in $Recall@100$ on the Touche2020 dataset, where the difference is minimal (0.4 difference). This consistent performance highlights the robustness of GOLFer in enhancing sparse retrieval tasks.
For dense retrieval using Contriever, GOLFer consistently surpasses other query expansion approaches across all low-resource datasets and metrics. Specifically, it shows substantial improvements in $nDCG@10$ and $Recall@100$ on datasets like NQ, FiQA-2018, and DBPedia.
These results demonstrate the effectiveness of GOLFer in enhancing query expansion performance, significantly contributing to the improvement of retrieval tasks.

\section{Analysis}\label{sec:05}

\paragraph{Ablation Study} To better understand the utility of GOLFer, we use Aggretriever\_v2 as a backbone model to conduct various experiments on the TREC DL 19/20 datasets, analyzing the impact and effectiveness of each component within this architecture as follows:

\textbf{Necessity of Individual Components}: We establish two variants to investigate the necessity of each component: a) \textbf{w/ Filter Only}: Our proposed framework with only the hallucination filter module. b) \textbf{w/ Combiner Only}: Our proposed framework with only the document combiner module.

\begin{table}[!ht]\small
    \centering
    \begin{tabular}{l|ccc}
    \hline
        \multirow{2}{*}{Model} & \multicolumn{3}{c}{TREC DL 19} \\ 
         & MAP & nDCG@10 & R@1K \\  \hline
        Aggretriever\_coCondenser & 43.5  & 68.4  & 80.8   \\ 
         ~~~w/ filter only & 47.3  & 68.6  & 86.0   \\ 
         ~~~w/ combiner only & 47.9  & 72.1  & 85.9  \\ 
         ~~~w/ filter + combiner & 48.4  & 73.0  & 86.4   \\ \hline
        ~ & \multicolumn{3}{c}{TREC DL 20} \\ 
        Aggretriever\_coCondenser & 47.1  & 69.7  & 85.6   \\ 
         ~~~w/ filter only & 45.7  & 64.6  & 87.4   \\ 
         ~~~w/ combiner only & 48.9  & 71.9  & 87.9   \\ 
         ~~~w/ filter + combiner & 49.0  & 72.2  & 88.3  \\ \hline
    \end{tabular}
\caption{Ablation results of GOLFer on TREC DL 19/20}
\label{tab6}
\end{table}

From Tabs \ref{tab6}, the following conclusions can be drawn: a) The performance of GOLFer on the TREC DL 19/20 datasets surpasses these variants lacking components, affirming the effectiveness and necessity of both the hallucination filter module and the document combiner module. The hallucination filter module reduces the degree of hallucination in smaller LM-generated passages, while the document combiner module balances the influence of the original query and the hypothetical document. These modules function independently yet complement each other, amplifying the performance of smaller LM-based query expansion. b) Among the different variants, the variant w/ Combiner Only shows high performance, highlighting the critical role of balancing the influence of the original query and the hypothetical document in enhancing query expansion. By further incorporating the hallucination filter module, irrelevant or erroneous information generated by smaller LMs is reduced, thus enhancing the overall performance of the GOLFer framework.

\paragraph{Compare to Large size Generative Models} In this experiment, we explore the potential of GOLFer using a smaller LM by comparing it with existing dominant query expansion methods with LLMs. Previous studies have shown that the scale of the generative LLM significantly impacts the quality of query expansion \cite{wang2023query2doc, gao2022precise}. We compared our performance to HyDE in BEIR datasets. It is important to note that the hypothesis documents for GOLFer are generated using LLaMA-3-8B-Instruct, while those for Hyde are generated by GPT-4o. 

\begin{table}[htbp]\small
\centering
\begin{tabular}{@{}lcc@{}}
\toprule
\multicolumn{1}{c}{}           & \multicolumn{1}{l}{GPT-4o} & \multicolumn{1}{l}{LLaMA-3 (8B)} \\
\multicolumn{1}{c}{}           & \multicolumn{2}{c}{w/ Contriever}                                 \\ \cmidrule(l){2-3} 
\multicolumn{1}{c}{}           & \multicolumn{1}{l}{w/ HyDE}      & \multicolumn{1}{l}{w/ GOLFer}  \\ \midrule
\multicolumn{1}{l|}{Scifact}   & 69.2                             & \textbf{69.4}                  \\
\multicolumn{1}{l|}{TREC-NEWS} & \textbf{44}                      & 41.1                           \\
\multicolumn{1}{l|}{FiQA}      & 27.6                             & \textbf{28.1}                  \\
\multicolumn{1}{l|}{DBPedia}   & \textbf{37.1}                    & 35.7                  \\ \bottomrule
\end{tabular}
\caption{Results for effect of different combination of instruction LMs and query expansion approaches. Hypothesis documents for GOLFer are generated using LLaMA-3-8B-Instruct, while those for Hyde are generated by GPT-4o. Best systems are marked \textbf{bold.}}
\label{tab4}
\end{table}

As shown in Table \ref{tab4}, GOLFer with LLaMA-3-8B outperforms HyDE with GPT-4o on the SciFact and FiQA datasets in terms of nDCG@10, although it falls behind on the TREC-NEWS and DBPedia datasets. Those results show that GOLFer with smaller LMs is competitive with, and sometimes outperforms, other query expansion methods with LLMs across various low-resource retrieval tasks. GOLFer is potential as a viable alternative to LLM-based query expansion methods in information retrieval.

\section{Conclusion }\label{sec:06}
In this work, we introduce GOLFer, a novel method designed to leverage smaller open-source LMs for query expansion, aiming to enhance both sparse and dense retrieval systems. The core idea is to distill the smaller LM outputs through effective hallucination detection and mitigation techniques. GOLFer identifies and filters out non-factual and inconsistent sentences in smaller LM-Generated documents, ensuring that only reliable documents are used as hypothetical documents for query expansion. The expanded query embeddings for information retrieval are then obtained by computing the dot product of the embedding vectors of the filtered hypothetical documents and the original query with a weight vector. Experimental evaluations demonstrate that the effectiveness of GOLFer in filtering and combining smaller LM-Generated texts contributes significantly to the improvement of query expansion performance in information retrieval. 

\section{Limitations }\label{sec:07}
We acknowledge several limitations in this paper. One significant limitation is the dependency on the self-attention mechanism of Transformer-based LLMs for evaluating factuality scores within the hallucination detection module. Although self-attention scores are available for all open-source LLMs, our method cannot be applied directly to certain APIs that do not offer access to these scores. Consequently, our future work will focus on developing alternative approaches to address this limitation.

\appendix
\section{Appendix}\label{A:01}

\subsection{Instructions}\label{A:0101}

\paragraph{TREC DL19} Instruction message = \textit{"Please write a passage to answer the question. [question\_text]"}.

\paragraph{TREC DL20} Instruction message = \textit{"Please write a passage to answer the question. [question\_text]"}.

\paragraph{MS MARCO dev} Instruction message = \textit{"Please write a passage to answer the question. [question\_text]"}.

\paragraph{NQ} Instruction message = \textit{"Please write a passage to answer the question. [question\_text]"}.

\paragraph{FiQA-2018} Instruction message = \textit{"Please write a financial article passage to answer the question. [question\_text]"}.

\paragraph{TREC\_COVID} Instruction message = \textit{"Please write a scientific paper passage to answer the question. [question\_text]"}.

\paragraph{Signal-1m} Instruction message = \textit{"Please write a passage to answer the question. [question\_text]"}.

\paragraph{TREC\_NEWS} Instruction message = \textit{"Please write a news passage about the topic. [question\_text]"}.

\paragraph{Robsut04} Instruction message = \textit{"Please write a news passage about the topic. [question\_text]"}.

\paragraph{Touche2020} Instruction message = \textit{"Please write a counter argument for the passage. [question\_text]"}.

\paragraph{CQADupStack} Instruction message = \textit{"Please write a passage to answer the question. [question\_text]"}.

\paragraph{DBPedia} Instruction message = \textit{"Please write a passage to answer the question. [question\_text]"}.

\paragraph{SciFact} Instruction message = \textit{"Please write a scientific paper passage to support/refute the claim. [question\_text]"}.

\end{document}